\documentclass{emulateapj}
\newcommand{\beq}{\begin{equation}}
\newcommand{\eeq}{\end{equation}}
\newcommand{\kms}{km s$^{-1}$~}
\newcommand{\rsol}{{\rm R}_\odot}
\newcommand{\msol}{{\rm M}_\odot}

\begin{document}

\shorttitle{Magnetic Braking in Ultracompact Binaries}
\shortauthors{Farmer \& Roelofs}

\title{Magnetic Braking in Ultracompact Binaries} 

\author{Alison J. Farmer \& Gijs H.\,A. Roelofs}

\affil{Harvard--Smithsonian Center for Astrophysics, 60 Garden Street, Cambridge, MA 02138, USA;\\\{afarmer,groelofs\}@cfa.harvard.edu}

\begin{abstract}
Angular momentum loss in ultracompact binaries, such as the AM Canum Venaticorum stars, is usually assumed to be due entirely to gravitational radiation. Motivated by the outflows observed in ultracompact binaries, we investigate whether magnetically coupled winds could in fact lead to substantial additional angular momentum losses.

We remark that the scaling relations often invoked for the relative importance of gravitational and magnetic braking do not apply, and instead use simple non-empirical expressions for the braking rates. In order to remove significant angular momentum, the wind must be tied to field lines anchored in one of the binary's component stars; uncertainties remain as to the driving mechanism for such a wind. In the case of white dwarf accretors, we find that magnetic braking can potentially remove angular momentum on comparable or even shorter timescales than gravitational waves over a large range in orbital period. We present such a solution for the 17-minute binary AM CVn itself which admits a cold white dwarf donor and requires that the accretor have surface field strength $\simeq 6 \times 10^4$ G. Such a field would not substantially disturb the accretion disk. Although the treatment in this paper is necessarily simplified, and many conditions must be met in order for a wind to operate as proposed, it is clear that magnetic braking cannot easily be ruled out as an important angular momentum sink.

We finish by highlighting observational tests that in the next few years will allow an assessment of the importance of magnetic braking.

\end{abstract}
\keywords{binaries: close --- gravitational waves --- stars: winds, outflows --- stars: magnetic fields --- stars: individual (AM Canum Venaticorum)}

\section{Introduction}

Magnetic braking, the removal of stellar angular momentum by a magnetically coupled wind, is held responsible for the observed spin-down of main-sequence stars with age \citep{Mes68}. It is questionable whether the empirical single-star braking laws \citep{Sku72,Smi79} apply when the star is a member of a close binary. Nevertheless, their simple translation into binary settings provides the standard explanation for the apparent angular momentum loss rates in low-mass X-ray binaries and cataclysmic variable stars \citep{Ver81,Spr83,Pac83,Rap83,Ver84}, which are 10--100 times larger than gravitational radiation can supply.

In this paper we investigate angular momentum shedding in ultra-compact binaries. We focus specifically on the AM CVn stars, in which both donor and accretor are white dwarfs (see \citealt{Nel05} for a review). Assuming cold white dwarf donors, several short-period AM CVn stars are overluminous \citep{Roe07b} compared with the accretion luminosities expected from gravitational-wave emission alone. 
One possible explanation is that the donor stars in AM CVns are hot and massive: the corresponding increase in system mass relative to the degenerate-donor scenario would boost the predicted gravitational-wave-driven luminosity to levels consistent with observations. However, some donors are then required to be extremely hot, most likely formerly helium-burning stars. Although there exists a theoretical AM CVn formation channel involving helium star donors \citep{Nel01a,Yun08}, there is no observational evidence for its operation. In particular, AM CVn stars appear to be nitrogen-rich rather than carbon-rich, arguing against helium burning in most of them \citep{Roe09}.

Here we explore the other possibility: that additional angular momentum loss occurs due to magnetic braking in an outflow, in analogy with X-ray binaries and cataclysmic variables.
The approach must differ from these cases involving main-sequence donor stars. There are no empirical data on single white dwarf spin-down rates. Clearly, the empirical main-sequence braking rates cannot be extrapolated to white dwarfs: in particular, magnetic fields of main-sequence stars are dynamo-generated, while those of white dwarfs are of fossil origin \citep[e.g.][]{Wic00} and are thus unaffected by rotation rate.  Furthermore, winds from white dwarfs are neither predicted nor observed (except in some of the very hottest white dwarfs; \citealt{Wer95}). Among ultracompact binaries with white dwarf donors and accretors, however, strong winds \emph{are} observed, in the form of P Cygni profiles of high-excitation spectral lines in the UV \citep{Wad07}. Such spectral signatures are commonly modeled with biconical outflows (e.g., \citealt{Lon02}) that are usually assumed to originate in the accretion disk. 

Unfortunately a disk wind, magnetized or not, is severely limited in the amount of angular momentum it can remove, because the disk itself possesses only a small quantity \citep[see e.g.][]{Kin95}. The angular momentum content of each binary component is much larger. If winds are to significantly influence the orbital evolution, they must be coupled to magnetic fields anchored in one of the binary's component stars.

The necessity of such fields is not a concern: magnetic fields are detected in many white dwarfs at the $\gtrsim10^4$\,G level \citep{Lie03}. The incidence of strong magnetic fields among white dwarfs in interacting binaries is higher than among single white dwarfs (e.g., \citealt{Tou08}), and there is evidence for a magnetic white dwarf in at least one ultracompact AM CVn binary \citep{Roe09}.

If the angular momentum carried by a magnetic stellar wind is ultimately to be drained from the binary orbit, the magnetic component's spin needs to be tidally or otherwise coupled to the orbit. The timescale for tidal spin--orbit synchronization is notoriously poorly known for white dwarfs due to the uncertain rate of viscous damping (e.g.\ \citealt{Ibe98,Mar04}) and the possibility of non-dissipative tidal coupling \citep{Rac07}. Several observational results suggest, however, that coupling may be sufficiently effective to produce synchronization of both white dwarfs in at least some systems \citep{Roe06,Roe07a}.

The conditions necessary for the operation of magnetic braking thus appear to be met. We adopt a first-principles approach to determine whether angular momentum shedding in an AM CVn magnetic wind could rival the inevitable loss due to gravitational radiation. It is tempting to dismiss this possibility by using the strong period dependence of gravitational radiation as proof of its dominance at very short periods over other types of angular momentum loss. The temptation must be resisted: the larger mass ratios typical for ultra-compact binaries may strongly diminish gravitational-wave emission, and the stronger surface magnetic fields common for compact stars boost their magnetic braking potential.

\section{A Simple Model}
\label{sec:simple}

We employ an elementary approach to estimate the angular momentum loss rate due to a magnetized wind, based on that of \cite{Web67}. Axial symmetry is adopted, all field lines are assumed open, and we focus on the equatorial plane. Caveats and exceptions will be considered in Section \ref{sec:caveats}. The wind corotates with the magnetized body at angular speed $\Omega$ out to the Alfv\'en radius $r_A$, which is where
\begin{equation}
\frac{B_p^2}{4 \pi \rho_w} = u^2 \,  ,
\label{eq:alf}
\end{equation}
in which $B_p$ is the poloidal field (equal to $B_r$ in the equatorial plane), $u$ is the wind velocity, and $\rho_w$ its density.
Conservation of mass,
\begin{equation}
\dot{M}_w = 4 \pi r^2 \rho_w u , 
\end{equation}
where $\dot M_w$ is the wind mass loss rate, and conservation of magnetic flux,
\begin{equation}
B_r r^2 = {\rm constant} ,
\end{equation}
can be combined with Eq.~(\ref{eq:alf}) to yield $r_A$:
\begin{equation}
r_A^2 = \frac{r_0^4 B_0^2}{u_A \dot{M}_w} \, .
\label{eq:alfven}
\end{equation}
where $r_0$ is the radius of the body emitting the wind, and $B_0$ is the magnetic field strength at the body's surface.  The wind velocity at the Alfven radius, $u_A$, is close to its terminal value \citep[see, e.g.,][]{Bel76}, so we adopt $u_A \simeq u_\infty$. The angular momentum loss rate is then
\begin{equation}
\dot J_\mathrm{mb} \simeq -\dot M_w \Omega r_A^2 \simeq  -\frac{\Omega r_0^4 B_0^2}{u_\infty} \, .
\label{eq:msw}
\end{equation}
This is to be compared with the angular momentum loss rate due to gravitational radiation,
\beq
\dot J_{\rm gw} = -\frac{32}{5} \frac{G^{7/3}}{c^5} \frac{M_1^2 M_2^2}{(M_1+M_2)^{2/3}} \Omega^{7/3} \, .
\eeq

\section{A specific example: AM CVn}
\label{sec:amcvn}

As a case in point, let us examine the prototypical interacting binary white dwarf, AM Canum Venaticorum. Observations reveal an orbital period of 1029\,s \citep{Ski99,Nel01b}, an accretion luminosity $L_{\rm acc} \simeq 2 \times 10^{34} {\rm~erg~s}^{-1}$ \citep{Roe07b} and the presence of an outflow with terminal velocity $u_\infty = 2-3\times 10^3$ \kms \citep{Wad07}.  The mass ratio $q \equiv M_2/M_1 = 0.18$, and there is evidence for synchronous rotation of both white dwarfs \citep{Roe06}.

Assuming that angular momentum is lost only via gravitational radiation, a total system mass $M \simeq 0.8 \msol$ \citep{Roe07b} is required. In order that the  $0.1 \msol$ donor fill its Roche lobe its radius must be $R_2 \simeq 0.046 \rsol$, implying a density much lower than a cold white dwarf of this mass. The donor in this case would need to be a semidegenerate helium star that has probably engaged in some core helium burning.

As described in the Introduction, we instead explore the scenario in which the Roche lobe-filling donor star is a cold white dwarf, and the angular momentum loss is supplemented by magnetic braking via a wind. We suppose the observed outflow to be this wind, and propose that it originates near the accretor, on field lines threading its surface. The total angular momentum loss rate is
\beq
\dot J = \dot J_{\rm gw} + \dot J_{\rm mb} \, ,
\eeq
and the disk accretion luminosity is related to $\dot J$ by the standard relation
\beq
L_{\rm acc} = \frac{\Delta \Phi}{2} \dot M= \frac{\Delta \Phi M_2}{(\zeta +5/3-2 q)} \frac{\dot J}{J_{\rm tot}} \, ,
\eeq
where $\zeta=d\ln R_2/d\ln M_2$. We approximate the potential drop $\Delta \Phi \simeq -G M_1/R_1$.
The cold white dwarf mass-radius relation for the donor star is taken to be
\begin{equation}
\frac{R_2}{R_\odot} = 0.0106-0.0064 \ln\left(\frac{M_2}{\msol}\right)+0.0015\left(\frac{M_2}{\msol}\right)^2 \, ,
\end{equation}
valid for low mass white dwarfs \citep{Nel01a}. For the higher-mass accreting white dwarf we use the fit of \citet{Ver88} and allow for a radius 5\% larger than the zero-temperature value,  as in \cite{Roe07b}, due to accretion heating. 

\begin{figure}
\begin{center}
\includegraphics[width=3.2in]{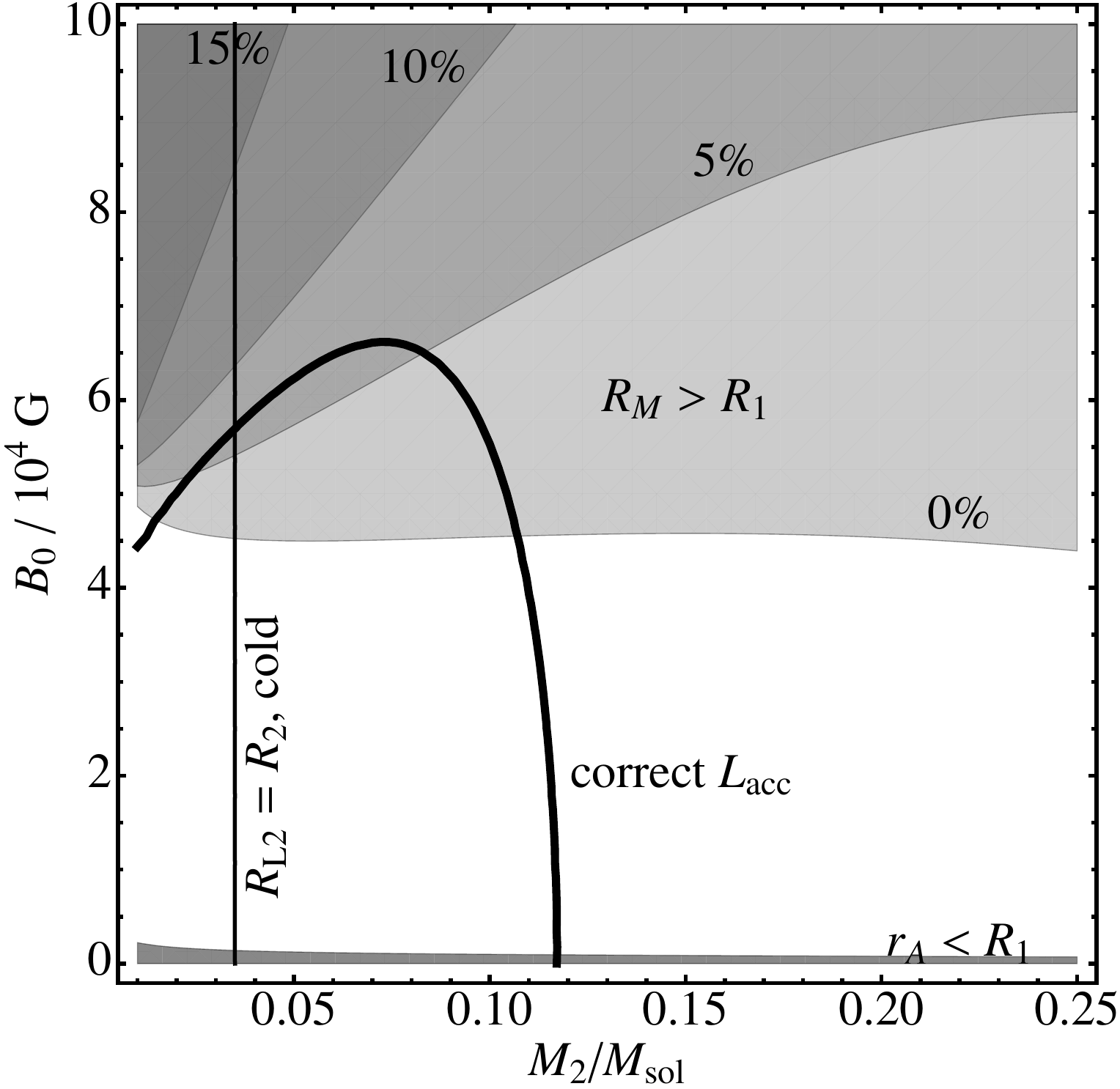}
\caption{Parameter space of solutions for AM CVn in terms of donor mass and accretor surface magnetic field strength. Thick solid line: systems with observed accretion luminosity; thin solid line: donor mass for which a cold white dwarf fills its Roche lobe. Shaded areas denote $\tau = (r_m-R_1)/(R_{L1}-R_1)>0, 0.05, 0.10, 0.15$. Note the zero magnetic field solution at $M_2 \simeq 0.1 \msol$ as described in the text. The solution in which the donor is a cold white dwarf requires $B_0 \simeq 6 \times 10^4$ G; for this solution, disk truncation occurs close to the surface of the accretor where it may go unobserved. Also shown is the region in which $r_A < R_1$ (for $\dot M_w = 0.01 \dot M$), where magnetic braking cannot occur.}
\label{fig:amcvn}
\end{center}
\end{figure}

Plotted as the thick line on Figure \ref{fig:amcvn} is the locus of donor mass and accretor surface magnetic field strength that together yield the observed accretion luminosity. Only a donor of mass $M_2 \simeq 0.035 \msol$ is permitted if it is to be cold and fill its Roche lobe (the thin vertical line). This solution requires a fairly modest accretor surface field $B_0 \simeq 6 \times 10^4$ G.

As in most AM CVn systems, the accretion stream in AM CVn does not seem to be magnetically channeled: the disk extends to near the surface of the accretor. This places an upper limit on the accretor's magnetic field strength. For a dipole field the disk truncation distance $r_m$ is
\begin{equation}
r_m \simeq 0.5 R_1 \left(\frac{B_0^4 R_1^5}{2 G M_1 \dot M^2}\right)^{1/7}
\label{eq:truncation}
\end{equation}
from the center of the accreting white dwarf \citep{Gho79}. The above can also be written in terms of the accretion luminosity:
\begin{equation}
r_m \simeq 0.5 R_1 \left(\frac{B_0^4 R_1^3 GM_1}{8 L_{\rm acc}^2}\right)^{1/7} \, .
\end{equation}
The absence of truncation implies $r_m\leq R_1$. We express the degree of truncation as a fraction $\tau$ of $(R_{L1}-R_1)$, where $R_{L1}$ is the distance from the center of the accretor to the inner Lagrange point L1. Given observational uncertainties, a slightly truncated disk, say $\tau = (r_m-R_1)/(R_{L1}-R_1) \lesssim 10\%$, would be indistinguishable from zero truncation. On Figure \ref{fig:amcvn} are displayed regions with truncation fractions $\tau <$5\%, $<$10\% and $<$15\%. The cold white dwarf solution for AM CVn yields $\tau \simeq 6\%$, where disk truncation may have gone undetected. We also note that the Eq. (\ref{eq:truncation}) is rather approximate. (See Section \ref{sec:caveats} for a discussion of the magnetic topology).

The zero-temperature donor star solution entails a total system mass $M=0.23$\,M$_\odot$, lower than predicted for AM CVn stars evolving under conservative mass transfer \citep{Nel01a}. More typical (according to \citealt{Nel01a}) system masses are obtained by assuming that the low-mass white dwarf donor has a realistic amount of remnant heat; see e.g.\ \citet{Del07}. A second possibility is a smaller mass ratio. Based on numerical accretion disk models, \citet{Woo09} argue that $q=0.06-0.08$ for AM CVn, rather than $q=0.18$ as derived by \citet{Roe06}. A cold donor would then imply a `typical' $M\simeq0.55$\,M$_\odot$, but the field required is slightly larger ($B_0 \simeq 8 \times 10^4$\,G) and the increased truncation ($\tau \simeq 13\%$) could be more difficult to reconcile with observations.

It appears that we can account for the observations of AM CVn while retaining the notion of the donor as a cold white dwarf. It is remarkable that the required accretor field strength is a plausible one, since 
there is no a priori reason for this to be so. In Section \ref{sec:caveats} we discuss the requirements and limitations of the solution.

\section{General Case}
Encouraged by the results of the previous section, we now turn to the general AM CVn system. We suppose that a typical system possesses an outflow from the accretor with velocity $u = 2000 {\rm~km~s}^{-1}$.

\begin{figure}
\begin{center}
\includegraphics[width=3.2in]{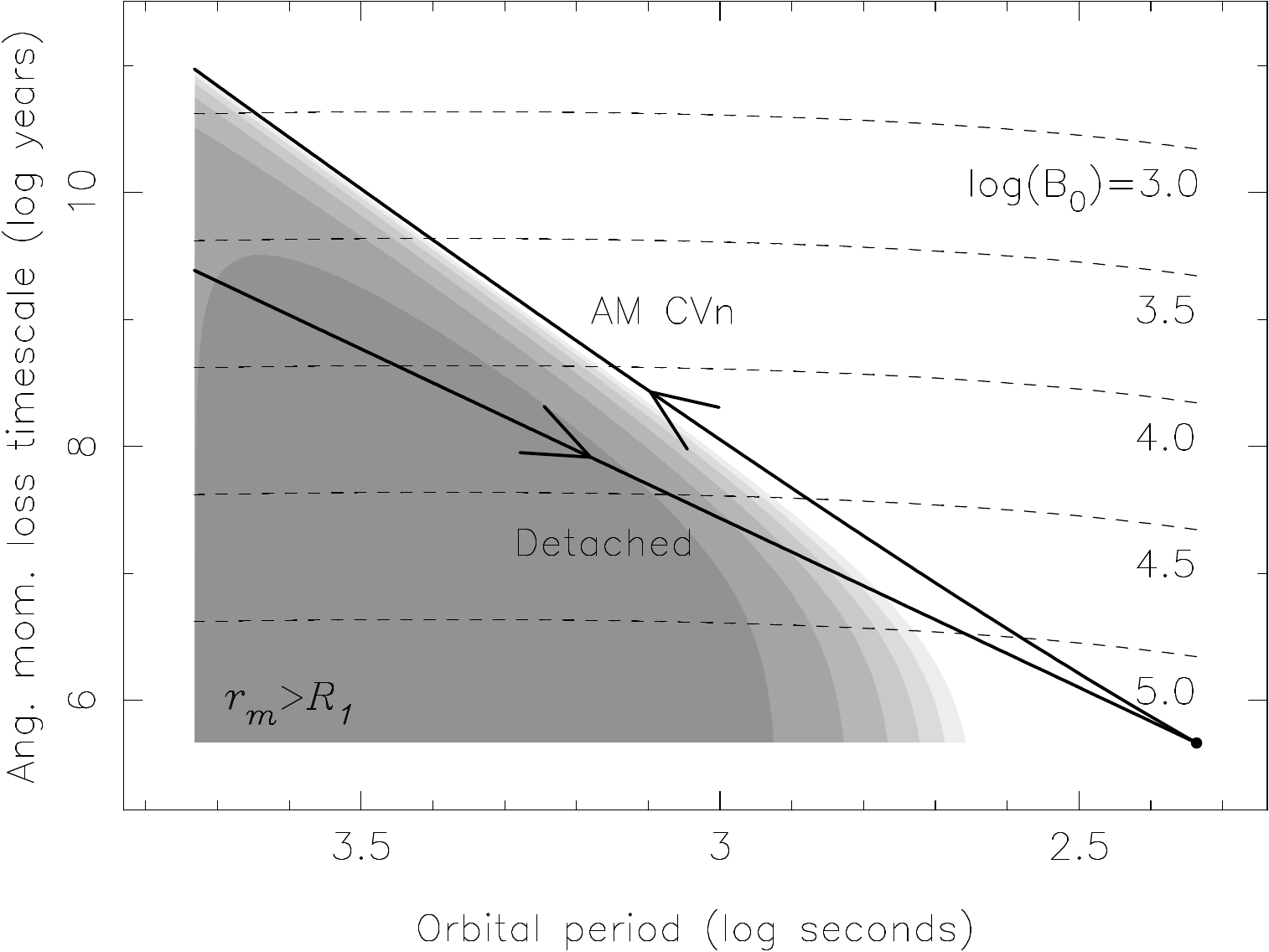}
\caption{Angular momentum loss timescales $(\equiv J_\mathrm{orb}/\dot J)$ due to gravitational radiation (solid line) and due to a magnetic wind (Eq.~(\ref{eq:msw}), dashed lines, AM CVn phase only), for various magnetic field strengths $B_0$\,(G) measured at the accretor before mass transfer starts. Shown is a `typical' $0.4+0.2\,M_\odot$ binary white dwarf. Grey shades indicate the amount of disk truncation $\tau$ in steps of 1\% of the distance between $R_1$ and $R_{L1}$, reaching a maximum of just under 7\%.}
\label{fig:porbdist}
\end{center}
\end{figure}

Figure \ref{fig:porbdist} compares the timescales for angular momentum loss, $J_\mathrm{orb}/\dot J$, via gravitational radiation and magnetic braking, during the life of a `typical' $0.4 + 0.2\,\msol$ system \citep{Nel01a}. The donor is a cold white dwarf, mass transfer is approximated as conservative and the magnetic flux $B_0 R^2_1$ through the accretor's surface is held constant.

Initially detached, gravitational radiation alone drives the binary to shorter orbital periods. Roche-lobe overflow commences at the minimum orbital period, after which we calculate the timescale for angular momentum loss due to the magnetically coupled wind of Eq.~(\ref{eq:msw}). Near period-minimum, fairly substantial surface fields of a few hundred kG would be required for magnetic braking to rival the gravitational-wave losses. As the binary evolves back to longer orbital periods during its semi-detached (AM CVn) phase, the required field strength decreases sharply. Without magnetic braking, the typical present-day AM CVn system lies where $J_{\rm orb}/\dot J_{\rm gw} \gtrsim$ 1 Gyr; a modest 10 kG surface field can then cause magnetic braking to dominate the angular momentum loss. At an orbital period of 1 hour, a mere 2\,kG field may suffice.

If the angular momentum loss rate is at all affected by magnetic braking, a binary will not follow the gravitational-wave driven AM CVn track in Fig.\ \ref{fig:porbdist}, but will evolve more quickly to longer orbital periods, along a path positioned below the gravitational-wave track. Once magnetic braking becomes dominant, the binary will evolve roughly parallel to the dashed lines in Fig.\ \ref{fig:porbdist}. The inner edge of the disk will then start to move away from the accretor. As shown in Fig.\ \ref{fig:porbdist}, however, disk truncation quickly reaches a maximum of $\tau \sim$7\%. 

\begin{figure}
\begin{center}
\includegraphics[width=3.2in]{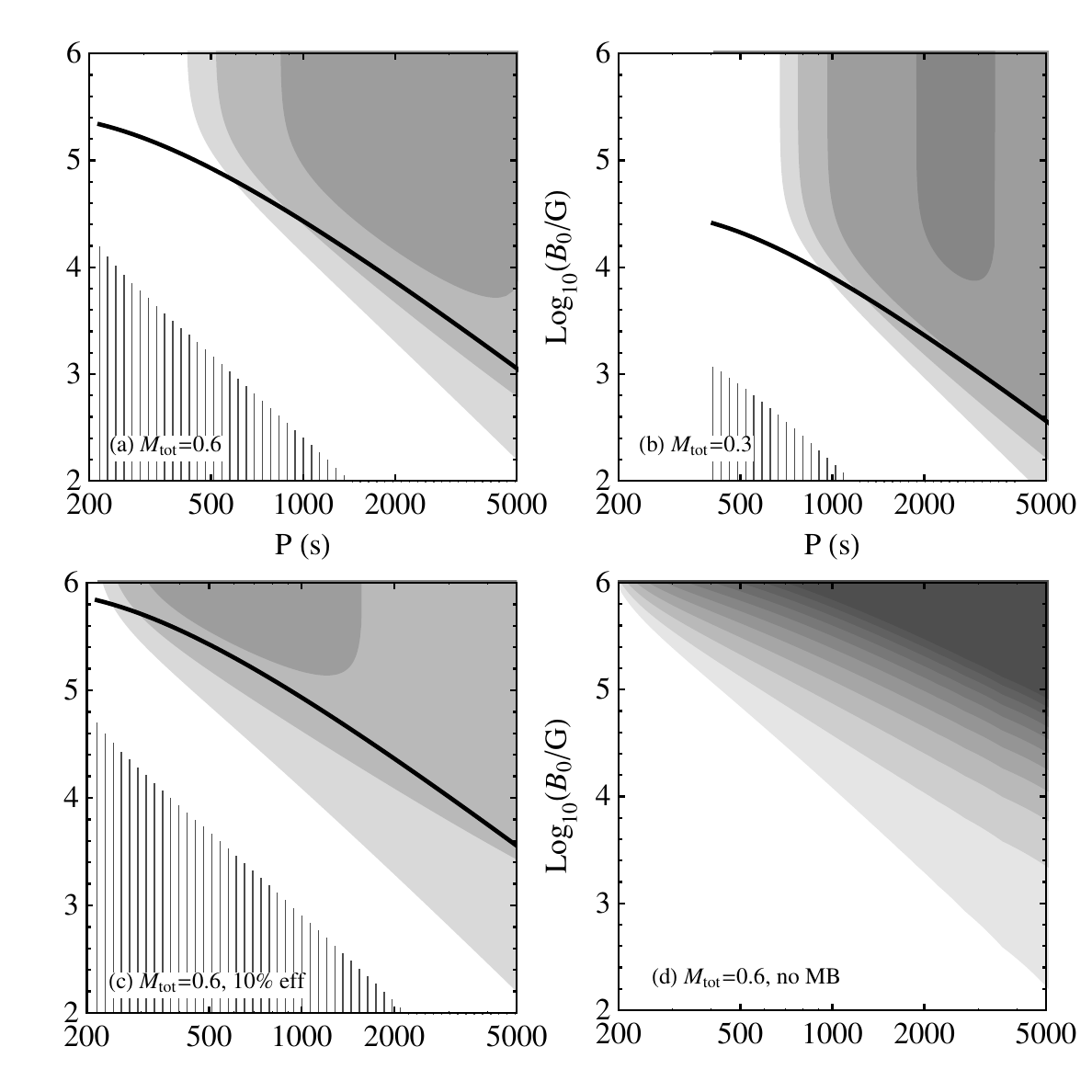}
\caption{Magnetic braking domination and disk truncation as a function of orbital period and surface magnetic field strength. (a) Total system mass $M_{\rm tot} = 0.6 {\rm M}_\odot$. Thick solid line  is the locus of $\dot J_{\rm gw} = \dot J_{\rm mb}$: above this line, magnetic braking dominates angular momentum loss. Solid shaded regions correspond to disk truncation from Eq.~(\ref{eq:truncation}), in steps of $\Delta \tau = 3\%$. Vertical hatched region is where $r_A < R_1$, for $\dot M_w = 0.01 \dot M$; in this region, magnetic braking cannot occur. (b) As (a), but for total system mass $0.3 {\rm M}_\odot$. (c) As (a), but with magnetic braking at 10\% efficiency; $r_A$ is assumed to scale with the square root of the efficiency. Disk truncation shading is now in steps of $\Delta \tau = 10\%$. (d) As (a), but without any magnetic braking; disk truncation shading is in steps of $\Delta \tau = 10\%$.}
\label{fig:4p}
\end{center}
\end{figure}

Figures \ref{fig:4p}a and \ref{fig:4p}b further illustrate this result: for two total system masses, they show the degree of disk truncation as a function of period and surface magnetic field strength. Where magnetic braking dominates, $r_m$ and thus $\tau$ are independent of $B_0$. The weak period dependence of 
$\tau$ results from the relatively gentle variation of $\dot J_{\rm mb}$ with period (cf. $\dot J_{\rm gw}$). Combined, these yield the broad maxima in truncation seen in Figures \ref{fig:4p}a and \ref{fig:4p}b.

As mentioned in Section \ref{sec:amcvn}, it is noteworthy that the (optimistic; see Section \ref{sec:caveats}) prediction of Eq.~(\ref{eq:msw}) should imply magnetic braking domination at field strengths observed routinely in white dwarfs. That this should lead to minimally truncated disks is also remarkable.

\section{Observations}
\label{sec:observations}

\begin{figure}
\centering
\includegraphics[width=3.2in]{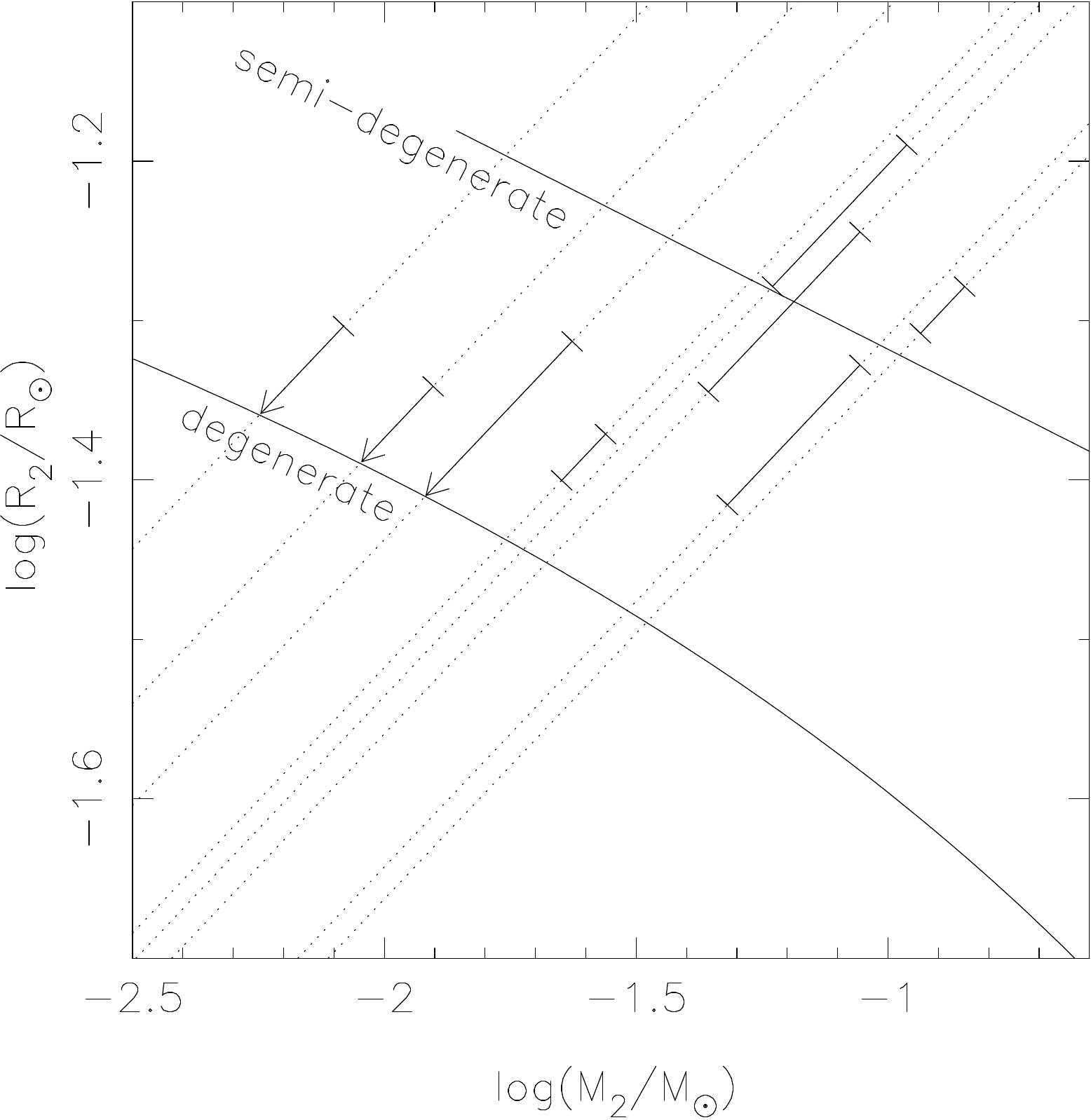}
\caption{Observational constraints on the masses and radii of the donors in 8 AM CVn stars, \emph{assuming the accretion luminosity is entirely gravitational-wave driven}. Luminosities of the 4 shortest-period systems (from the right: AM CVn, HP Lib, CR Boo, V803 Cen) are disk-dominated; those of the 4 longest-period systems (from the left: V396 Hya, GP Com, SDSS\,J1240, SDSS\,J0926) are accretor-dominated. Also shown are donor star evolutionary tracks for zero-temperature (degenerate) white dwarfs and `typical' semi-degenerate helium stars, the latter from \citet{Yun08}. Data collected and derived from \citealt{Roe05,Roe07b,Roe07c,Tho08}; Marsh et al.\ in preparation; Steeghs et al.\ in preparation.}
\label{fig:massesradii}
\end{figure}

The main driver for our investigation into magnetic braking was the apparent overluminosity of the short-period ($P\lesssim 30$ min) AM CVn stars. Interestingly, the observations at longer periods  are compatible with angular momentum loss via gravitational radiation alone, with fairly degenerate donors. Fig.\ \ref{fig:massesradii} collects the observational data (from \citealt{Roe06}, with more recent data points added). Of particular interest is the unique eclipsing AM CVn star, SDSS\,J0926+3624, with an orbital period of 28 minutes, for which the component masses have been measured from eclipse timing rather than inferred from the star's luminosity (Marsh et al., in preparation). The donor star need be only slightly more massive than a zero-temperature white dwarf and appears to be quite compatible with expectations for a white dwarf \citep{Del07}.

At first glance, the evidence for magnetic braking only at short orbital periods seems counter to expectations: at shorter periods, larger magnetic field strengths are required if magnetic braking is to rival gravitational radiation in angular momentum loss (see Figure \ref{fig:4p}). However, it may be quite natural that magnetic braking should favor the short-period systems. As described in the Introduction, winds are not expected from typical non-interacting white dwarfs. Because AM CVn outflows are powered by accretion energy (an example of `CAML'; \citealt{Kin95}), the smaller accretion luminosities of the longer-period systems may well lead to reduced (or no) winds. Although Eq.~(\ref{eq:msw}) shows $\dot J_{\rm mb}$ to be independent of $\dot M_w$, a weaker wind may in reality mean a smaller fraction of open field lines, and thus decreased $\dot J_{\rm mb}$. In the limit $\dot M_w \rightarrow 0$, magnetic braking will surely cease, if only because the ISM will stop a sufficiently feeble wind before it reaches the Alfv\'en radius. It could thus be that at long periods the wind stops and the binary collapses back onto the gravitational-wave driven track of Figure \ref{fig:porbdist}.

This explanation has difficulties: if braking dominates in the closest systems, but switches off as the binary evolves to longer periods, then all long-period systems would have significantly truncated accretion disks. Most AM CVn stars have accretion disks that are not substantially truncated.  Figure \ref{fig:4p}c plots the result of reducing $\dot J_{\rm mb}$ by a factor 10. The degree of disk truncation for given $B_0$ greatly exceeds that of the efficient braking case in Figure \ref{fig:4p}a (the shading levels in Figure \ref{fig:4p}a are in steps of $\Delta \tau = 10\%$; in Figure \ref{fig:4p}a they are $\Delta \tau = 3\%$).  If braking ceases completely, $\tau$ will reach still larger values (Figure \ref{fig:4p}d).

A second possible explanation for the presence of magnetic braking in only the short-period systems is as follows. If braking remains effective at long periods, then the period evolution of long-period magnetic systems is much faster than the evolution of those driven by gravitational radiation alone. Fewer of the magnetic systems would be detected in surveys. The disks of the (non-magnetic) detected systems would naturally be intact. Because the system luminosity is dominated by the cooling accretor at long periods (shown by \citealt{Bil06} for gravitational-wave domination), the larger accretion luminosity of the magnetic systems would not translate into correspondingly higher detection rates. With better population statistics on short- and long-period systems, this can be tested observationally: towards longer periods there would be fewer systems than expected based on gravitational-wave driving alone.

The recently discovered system SDSS\,J0804$+$1616 ($P= 44.5$\,min) does appear to have a strongly truncated disk \citep{Roe09}. This requires a magnetic field of order $\sim$100\,kG if the mass transfer rate is set by gravitational waves alone. This degree of disk truncation would not be possible if efficient magnetic braking dominated the angular momentum loss. Highly magnetic systems might be expected to have inefficient braking: very strong magnetic fields could produce a large dead zone enveloping the major wind-driving latitudes (see Sect.\ \ref{sec:caveats}).

\section{Discussion: Origin of the wind}
\label{sec:caveats}

In this section we address the limitations of our assumptions and consider alternatives.

Thus far we have assumed that magnetic braking occurs in an outflow of the type observed in AM CVn, and that the wind issues from the accretor. In CVs the wind is thought to originate in the inner accretion disk, because a radial wind is inconsistent with observations of line profiles \citep{Mau87,Fro05,Pro05}. However, even a modest accretor magnetic dipole would entrain such a disk wind and channel it on to the star. The field strength required is of order that which yields $r_A \gtrsim R_1$; this quantity is plotted on Figure \ref{fig:4p}a, assuming $\dot M_w$ is 1\% of the mass transfer rate. This field strength is always less than that at which braking dominates purely gravitational angular momentum loss. At $P=1000$s, only $B_0 \simeq 300$G is required to disrupt a disk wind. 

Another possibility is that much of the observed wind is from the disk, but that there exists a second component coming from the accretor. The independence of $\dot J_{\rm mb}$ from $\dot M_w$ means that even a minor wind from the star could shed significant angular momentum.

Implicit in the discussion so far has been the magnetic topology: the structure of the magnetic field around our putative wind-emitting accretor is necessarily speculative. Even in the absence of a wind, the structure is uncertain: field lines may be `inflated' \citep{Lov95}, depending on the relative magnitudes of viscosity and magnetic diffusivity in the disk. For simplicity we have assumed that the field lines are not disrupted by the disk flow. Magnetic field `burial' by accretion \citep{Cum02} is ignored.

The treatment in Section \ref{sec:simple} assumed a uniform radial wind following a split monopole field. In practice \citep[e.g.][]{Cam97} the field lines at lowest latitude are expected to be closed in a `dead zone'; the size $\bar r$ of the zone depends on, among other factors, the temperature of the wind. We employed a dipole approximation when calculating the `truncation radius' in Section \ref{sec:amcvn}, because close to the star in the equatorial plane, the field could be close to dipolar. The dead zone radius may be significantly smaller than $r_A$ \citep{Cam97}. Because only open field lines can participate in magnetic braking, the braking rate is diminished by a factor $(R_1/\bar r)^2$. If the wind is emitted from only a fraction of the open field lines, the braking rate is further decreased. Eq.~(\ref{eq:msw}) represents the theoretical maximum magnetic braking rate; in Figure \ref{fig:4p}c we illustrate the effects of reduced braking efficiency.

The predicted braking rate of Eq.~(\ref{eq:msw}) is independent of the wind mass loss rate, and by assumption therefore the mass transfer rate between the binary components. In reality, as discussed in Section \ref{sec:observations}, some dependence on mass transfer rate may be able to account for the `necessity' of magnetic braking only at short periods. Some variation is expected once such factors as dead zone radius and wind driving mechanisms are taken into account. Provided the dependence on $\dot M$ is sufficiently weak, mass transfer should not be destabilized \citep[e.g.][]{Kin95}.

A major question is the wind driving mechanism. The observed wind from AM CVn moves at close to the escape velocity from a white dwarf accretor, which is $v_e \simeq 2-3 \times 10^8~{\rm cm~s}^{-1}$ for a white dwarf of mass $M_1\simeq0.3\,M_\odot$. This suggests thermal driving as the main component:  centrifugally-dominated acceleration does not preferentially yield terminal velocities close to the escape velocity \citep{Bel76}.

Energetically there is no difficulty in driving a wind from the accretor, since only a small fraction of the total mass transferred is observed in the wind (\citealt{Fro05}, for CVs). Thermal driving requires a sound speed $a_s \simeq v_{\rm esc}$ near the launch site; the energy required to heat the gas is easily exceeded by that released by accretion in the boundary layer: $L \sim \dot M v_{\rm esc}^2/4 \gg \dot M_w a_s^2$. Similar reasoning was discussed in the context of protostars by \cite{Mat05}. 

A larger issue is how the hot material ends up on open field lines threading the accretor. It is clear that open field lines cannot thread both disk/boundary layer and star. Spreading of hot accreted material to higher latitudes is possible, but the spreading angle is probably much less than 90 degrees (\citealt{Pir04}; but see \citealt{Fis06}). An alternative heating source at high latitudes is irradiation by the accretion disk itself. Sufficient energy is available, but detailed calculations would be necessary in order to determine whether such heating would launch a wind.

In other binary systems in which magnetic braking is invoked, the wind emanates from the donor star, which in those cases is expected also to emit a wind while isolated. Cold white dwarfs with neutral atmospheres, however, do not independently emit plasma winds. Irradiation from the accretion flow could in principle ionize and heat the donor atmosphere enough to drive a wind \citep[e.g.][for low-mass X-ray binaries]{Tav93}. Such an outflow would be preferentially emitted from the side facing the center of mass of the system and would thus be inefficient in removing angular momentum.  A donor wind is, however, emitted from a star whose spin is concretely expected to be synchronized with the orbit. The magnetic field of such a stripped donor star is unknown, but for given $B_0$, its larger radius would boost the braking rate. Disk truncation might also be reduced. At long orbital periods ($\simeq$40\,min) the donor stars are predicted to become fully convective \citep{Del07} which could naturally lead to a change in magnetic field and thereby magnetic braking efficiency.

We have assumed throughout that inferred accretion rates from observations can be directly compared with theoretically determined ones. Prolonged departures from secular $\dot M$ values of the type discussed by \citet{Rit00} cannot be excluded, but are not expected for degenerate donors, whose radii are little changed by irradiation.

\section{Conclusions}

We have demonstrated that the apparent overluminosity of AM CVn, relative to the accretion luminosity expected due to gravitational radiation from two degenerate white dwarfs, can be explained if the observed outflow from AM CVn is coupled to the magnetic field of the accretor. Using a simplified wind model, the required field strength at the accretor's surface is a rather modest $B_0 \simeq 6 \times 10^4$ G. Such a field would not produce substantial truncation of the inner accretion disk, and would thus remain consistent with the disk observed. The approach in this paper is necessarily approximate, and many conditions must be met for a wind to operate as proposed, but our results demonstrate that magnetic braking cannot easily be ruled out as an important angular momentum sink.

Generalizing to the population of AM CVn stars, magnetic braking seems capable of rivaling gravitational radiation as the agent of angular momentum loss at most orbital periods. If magnetic braking operates at long periods, even inefficiently, there will be a natural observational bias against systems with strongly truncated accretion disks since they evolve much faster, which would explain why most long-period systems have disks that are largely intact. This should become apparent from the observed orbital period distribution of AM CVn stars once the population statistics improve.

Another promising test for angular momentum losses additional to gravitational radiation may be provided by the recently discovered eclipsing AM CVn star, SDSS\,J0926+3624 \citep{And05}. Eclipse timing will allow accurate measurements of both the stellar masses and the orbital period derivative (provided transient phenomena are unimportant); within a few years it can be checked whether these match with gravitational-wave radiation.

\acknowledgments

GHAR is supported by NWO Rubicon grant 680.50.0610 to G.H.A. Roelofs. AJF thanks Gijs Nelemans for helpful discussions and thanks everyone for the good times.

\end{document}